\documentstyle[12pt,epsfig]{article}

\newcommand{\be}{\begin{equation}}
\newcommand{\ee}{\end{equation}}

 1
\font\elevenrm=cmr10 scaled\magstep 1
 1

\def\refe{\hang\noindent}
\def\ltsima{$\; \buildrel < \over \sim \;$}
\def\simlt{\lower.5ex\hbox{\ltsima}}            
\def\gtsima{$\; \buildrel > \over \sim \;$}
\def\simgt{\lower.5ex\hbox{\gtsima}}            

\def\AAP{Astronomy and Astrophysics}
\def\AAS{Astronomy and Astrophysics Supplement Series}
\def\AJ{Astronomical Journal}
\def\APP{Astroparticle Physics} 
\def\ANNREV{Annual Review of Astronomy and Astrophysics}
\def\APJ{Astrophysical Journal}
\def\APJL{Astrophysical Journal (Letters)}
\def\APJS{Astrophysical Journal Supplement Series}

\def\MN{Monthly Notices of the Royal Astronomical Society}
\def\PASP{Publications of the Astronomical Society of the Pacific}
\def\NAT{Nature}
\def\SAIT{Memorie della Societ\`a Astronomica Italiana} 

\begin{document}
\vspace*{1.8cm}
  \centerline{\bf HIGH ENERGY EMISSION FROM AGN AND UNIFIED 
SCHEMES\footnotemark}
\vspace{1cm}
\footnotetext{Invited Review Talk at the Vulcano Workshop {\it Frontier 
Objects in Astrophysics and Particle Physics}, Vulcano, Italy, May 1998}
  \centerline{PAOLO PADOVANI}
\vspace{1.4cm}
  \centerline{Space Telescope Science Institute}
  \centerline{\elevenrm 3700 San Martin Drive, Baltimore, MD. 21218, USA}
  \centerline{\elevenrm E-mail: padovani@stsci.edu}
\vspace{0.3cm}
  \centerline{Affiliated to the Astrophysics Division, Space Science 
Department, European Space Agency}
\vspace{0.3cm}
  \centerline{On leave from Dipartimento di Fisica, II Universit\`a di Roma,
Italy}
\vspace{3cm}
\begin{abstract}
Active Galactic Nuclei (AGN) are now known to be strong $\gamma$-ray
emitters. After briefly describing AGN classification and the main ideas
behind unified schemes, I summarize the main properties of blazars (that is BL
Lacs and flat-spectrum radio quasars) and their connection with relativistic
beaming. Finally, I address the question of why blazars, despite being extreme
and very rare objects, are the only AGN detected at high ($E > 100$ MeV)
energies, and touch upon the relevance of TeV astronomy for AGN research.
\end{abstract}
\vspace{2.0cm}

\section{Active Galactic Nuclei}\label{sec:agn}

Active Galactic Nuclei (AGN) are extragalactic sources, in many cases clearly
associated with nuclei of galaxies (although in the most distant objects the
host galaxy is too faint to be seen), whose emission is dominated by
non-stellar processes in some waveband(s) (typically but not exclusively the
optical). One important feature of AGN is the fact that their emission covers
the whole electromagnetic spectrum, from the radio to the $\gamma$-ray band,
sometimes over almost 20 orders of magnitude in frequency.

It is now well established that AGN are strong $\gamma$-ray ($E > 100$ MeV)
emitters. To be more specific: 1. at least 40\% of all EGRET sources are AGN
(Thompson et al. 1995, 1996; some more AGN are certainly present amongst the
still unidentified sources) and these make up almost 100\% of all
extragalactic sources (the only exceptions being the Large Magellanic Cloud
and possibly Centaurus A); 2. {\it all} detected AGN are blazars, that is BL
Lacertae objects (BL Lacs) or flat-spectrum radio quasars
(FSRQ).\footnote{The term ``blazar'' is here given a wider meaning than the
one sometimes implied, which is restricted to highly polarized quasars (HPQ)
and/or optically violently variable (OVV) quasars. The reason is that there is
increasing evidence that these categories and the flat-spectrum radio quasars,
which reflect different empirical definitions, refer to more or less the same
class of sources. That is, the majority of flat-spectrum radio quasars tend to
show rapid variability and high polarization.} To appreciate the relevance of
the latter point, we will first have to tackle the subject of AGN
classification.

\subsection{The AGN Zoo}\label{sec:class}

The large number of classes and subclasses appearing in AGN literature could
disorientate physicists or even astronomers working in other fields. A
simplified classification, however, can be made based on only two parameters,
that is radio-loudness and the width of the emission lines, as summarized in
Table 1 (see, e.g., Urry and Padovani 1995). 

\begin{table}[t]
\vspace{0.4cm}
\begin{center}
\begin{tabular}{|l|l|l|l|}         
\hline
{\bf Radio Loudness} & \multicolumn{3}{c|}{\bf Optical Emission Line
Properties} \\ \hline 

 & {\bf Type 2} (Narrow Line) & {\bf Type 1} (Broad Line) & {\bf Type 0} 
(Unusual) \\ 
 & & & \\
Radio-quiet: & Seyfert 2 & Seyfert 1 &  \\
 & & & \\
             &           & QSO       &  \\ 
 & & & \\
\hline
 & & & \\
Radio-loud: & NLRG $\cases {{\rm FR~I} \cr ~ \cr {\rm FR~II} \cr}$ & 
BLRG & Blazars $\cases {{\rm BL~Lacs} \cr ~ \cr {\rm (FSRQ)} \cr}$ \\

            &    & SSRQ & \\
            &    & FSRQ & \\
 & & & \\
\hline
 & \multicolumn{3}{c|}{decreasing angle to the line of sight $\longrightarrow$} \\
\hline
\end{tabular}
  \caption{\em {AGN Taxonomy: A Simplified Scheme}}
\end{center}
\end{table}

Although it was the strong radio emission of some quasars that led to their
discovery about 35 years ago, it soon became evident that the majority of
quasars were actually radio-quiet, that is most of them were not detected by
the radio telescopes of the time. It then turned out that radio-quiet did not
mean radio-silent, that is even radio-quiet AGN can be detected in the radio
band. Why then the distinction? If one plots radio luminosity versus optical
luminosity for complete samples of optically selected sources, it looks like
there are two populations, the radio quiet one having, for the same optical
power, a radio power which is about $3 - 4$ orders of magnitudes smaller.
The distribution of the luminosity ratio $L_{\rm r}/L_{\rm opt}$ for complete
samples, including the upper limits on the radio luminosity, appears to be
bimodal, with a dividing line at a value $L_{\rm r}/L_{\rm opt} \sim 10$
(e.g., Stocke et al. 1992; both luminosities are in units of power/Hz).
It would therefore be better to call the two classes ``radio-strong'' and
``radio-weak'' but the original names are still used. Note that only about $10
- 15\%$ of AGN are radio-loud. 

The other main feature used in AGN classification is the width, in case they 
are present, or the absence, of emission lines. These are produced by the
recombination of ions of various elements (most notably H, He, C, N, O, Ne,
Mg, Fe). Their width is due to the Doppler effect thought to result from
more or less ordered motion around the central object. AGN are then divided in
Type 1 (broad-lined) and Type 2 (narrow-lined) objects according to their
line-widths, with 1000 km/s (full width half maximum) being the dividing
value. Some objects also exist with unusual emission line properties, such as
BL Lacs, which have very weak emission lines with typical equivalent widths (a
measure of the ratio between line and continuum luminosity) $< 5$ \AA. 

As illustrated in Table 1, we then have radio-quiet Type 2 and Type 1 AGN,
that is Seyfert 2 galaxies and Seyfert 1 galaxies/radio-quiet quasars (QSO)
respectively. Radio-loud Type 2 AGN are radio galaxies (sometimes also called
narrow-line radio galaxies [NLRG] to distinguish them from the broad-lined
ones), classified as Fanaroff-Riley (Fanaroff and Riley 1974) I and II (FR I
and II) according to their radio morphology (connected with their radio
power), while radio-loud Type 1 AGN are broad-line radio galaxies (BLRG) and
radio quasars.  Finally, radio-loud sources with very weak emission lines are
known as BL Lacertae objects, from the name of the class prototype, which was
originally presumed to be a variable star in the Lacerta constellation.

Concentrating on the radio-loud sources, to which most of this paper is
devoted, the BLRG are, at least in my view, simply local versions of radio
quasars where we can detect the host galaxy, as Seyfert 1 galaxies are local
versions of QSO (the possible reasons why we do not see the high-redshift
counterparts of Seyfert 2 galaxies are discussed by Padovani 1998). Radio
quasars are generally divided into steep-spectrum radio quasars (SSRQ) and
flat-spectrum radio quasars (FSRQ), according to the value of their radio
spectral index at a few GHz ($\alpha_{\rm r} = 0.5$ being usually taken as the
dividing line, with $f_{\nu} \propto \nu^{-\alpha}$). This distinction
reflects the size of the radio emitting region. In fact, radio emission in
these sources is explained in terms of synchrotron radiation (that is
radiation from relativistic particles moving in a magnetic field), which for
extended regions has a relatively steep spectrum ($\alpha_{\rm r} \sim 0.7$).
On the other hand, nuclear, compact emission has a flatter spectrum, thought
to be the result of the superposition of various self-absorbed components. The
flat radio spectrum then indicates that nuclear emission dominates over the
more extended emission, generally associated with the so-called
``radio-lobes.''  In fact, flat-spectrum quasars are generally core-dominated
in the radio band, that is emission from the core is much stronger than
emission from the extended regions, unlike for example SSRQ or narrow-line
radio galaxies which are both lobe-dominated. Note that even though FSRQ have
strong broad lines they are also included in the ``Type 0'' column in Table 1
because their multifrequency spectra are dominated by non-thermal emission as
in BL Lac objects.

\subsection{Unified Schemes}\label{sec:unif}

\begin{figure}[p]
\hskip 1.5truecm 
\psfig{figure=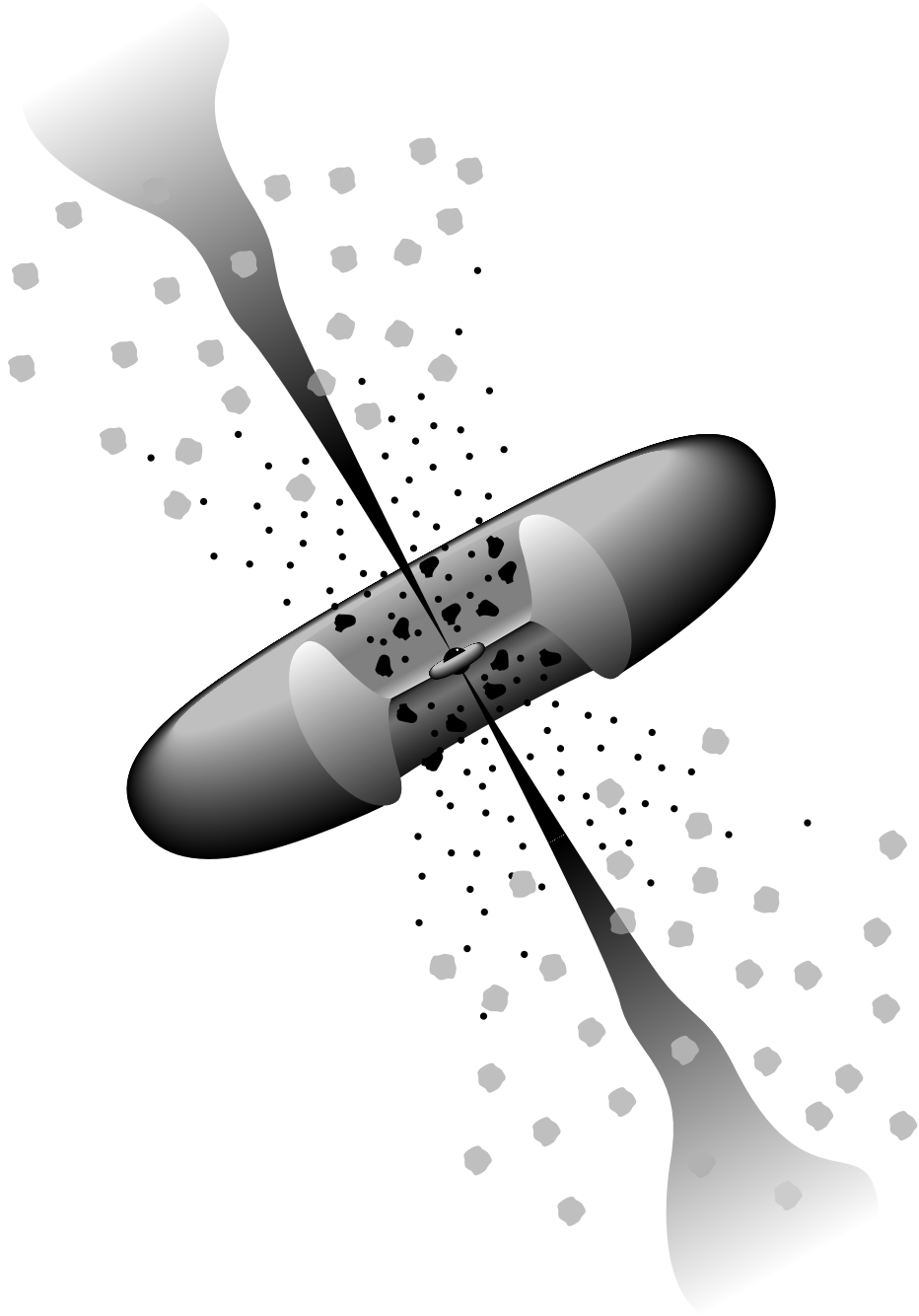,height=18truecm}
\caption{A schematic (and highly idealized) diagram of the current paradigm
for radio-loud AGN (not to scale). Surrounding the central black hole is a
luminous accretion disk.  Broad emission lines are produced in clouds (dark
spots) orbiting above the disk and perhaps by the disk itself. A thick dusty
torus (or warped disk) obscures the broad-line region from transverse lines of
sight; some continuum and broad-line emission can be scattered into those
lines of sight by hot electrons (black dots) that pervade the region. A hot
corona above the accretion disk may also play a role in producing the hard
X-ray continuum.  Narrow lines are produced in clouds (grey spots) much
farther from the central source. Radio jets, shown here as the diffuse jets
characteristic of low-luminosity, or FR~I-type, radio sources, emanate from
the region near the black hole, initially at relativistic speeds (Urry and
Padovani 1995; copyright Astronomical Society of the Pacific, reproduced with
permission).}
\label{fig:unif}
\end{figure}

All this might seem complicated, but in recent years we have developed a
consistent scenario which at least explains the Type 0/1/2 distinction. We
have in fact come to understand that some classes of apparently different (and
therefore classified under different names) AGN might actually be
intrinsically the same class of objects seen at different angles with the line
of sight (see for example Antonucci 1993 and Urry and Padovani 1995 and
references therein).

The main idea, based on various observations and summarized in Fig. 1, is that
emission in the inner parts of AGN is highly anisotropic. The current paradigm
for AGN includes a central engine, possibly a massive black hole, surrounded
by an accretion disk and by fast-moving clouds, probably under the influence
of the strong gravitational field, emitting Doppler-broadened lines. More
distant clouds emit narrower lines.  Absorbing material in some flattened
configuration (usually idealized as a toroidal shape) obscures the central
parts, so that for transverse lines of sight only the narrow-line emitting
clouds are seen (Type 2 AGN), whereas the near-infrared to soft-X-ray nuclear
continuum and broad-lines are visible only when viewed face-on (Type 1
AGN). In radio-loud objects we have the additional presence of a relativistic
jet, roughly perpendicular to the disk, which produces strong anisotropy and
amplification of the continuum emission (``relativistic beaming''), which I
discuss in more detail in Sect. \ref{subsec:prop}. For reasons still unclear,
BL Lac objects have extremely weak emission lines, and their continuum is
very strong and non-thermal (i.e., due to synchrotron and, at high
energies, inverse Compton emission or perhaps hadronic processes).

This axisymmetric model of AGN implies widely different observational
properties (and therefore classifications) at different aspect angles. Hence
the need for ``Unified Schemes'' which look at intrinsic, isotropic
properties, to unify fundamentally identical (but apparently different)
classes of AGN. Seyfert 2 galaxies have therefore been ``unified'' with
Seyfert 1 galaxies, whilst low-luminosity (FR I) and high-luminosity (FR II)
radio galaxies have been unified with BL Lacs and radio quasars respectively
(see Antonucci 1993 and Urry and Padovani 1995 and references
therein). In other words, BL Lacs are thought to be FR I radio galaxies with
their jets at relatively small ($\simlt 20 - 30^{\circ}$) angles w.r.t. the
line of sight. Similarly, we believe FSRQ to be FR II radio galaxies oriented
at small ($\simlt 15^{\circ}$) angles, while SSRQ should be at angles
in between those of FSRQ and FR II's ($15 \simlt \theta \simlt 40^{\circ}$).
Blazars are then a special class of AGN which we think have their jets 
practically oriented towards the observer. 

In general, different AGN components are important at different wavelengths.
Na\-me\-ly: 1. the jet emits non-thermal radiation, via electromagnetic
(synchrotron and inverse Compton) and perhaps hadronic processes, all the way
from the radio to the $\gamma$-ray band (Mastichiadis 1998; Dar 1998); 
2. the accretion disk probably emits thermal radiation, peaked in
optical/ultraviolet/soft-X-ray band; 3. the obscuring material (torus) will
emit predominantly in the infrared. These different components are apparent,
for example, in the multifrequency spectrum of 3C 273 (Lichti et al. 1995)
the first quasar to be discovered and one of the best studied.

At this point one might ask: what has all this to do with $\gamma$-ray
emission? The answer in the next section.  

\section{Blazars as $\gamma$-ray Sources}\label{sec:role}

According to Unified Schemes, blazars are that special class of radio-loud AGN
with their jets pointing more or less towards the observer, and therefore
constitute a relatively rare class of objects. Radio-loud AGN make up only
$\sim 10 - 15\%$ of all AGN (e.g., Kellermann et al. 1989), while a generous
upper limit to the fraction of blazars amongst radio sources is 50\% (as
inferred, for example, from the fraction of FSRQ and BL Lacs in the 1 Jy
catalogue [Stickel et al. 1994] which, being a high-frequency radio catalogue,
is biased towards flat-spectrum sources). It then follows that blazars make up
at most 5\% of all AGN, but more likely even less than that.

Mukherjee et al. (1997) have identified 51 high-confidence EGRET sources
(mainly from the Second EGRET catalogue; Thompson et al. 1995, 1996) with AGN,
all of them blazars. If the probability of detecting an AGN with EGRET were
independent of the class, then in this list we would expect at maximum 3
blazars, with most sources being associated with radio-quiet AGN. Instead, we
have 100\% blazars and 0\% other sources. In particular, no radio-quiet AGN
has been detected so far by EGRET. Note that blazar $\gamma$-ray luminosities
are in the range $10^{45} - 10^{49}$ erg/s (under the assumption of
isotropy) and in many cases the output in
$\gamma$-rays dominates the total (bolometric) luminosity.

To find out what is so special about blazars we need to have a closer look at 
their properties. 

\subsection{Blazar Properties and Relativistic Beaming}\label{subsec:prop}

The main properties of blazars can be summarized as follows:

\begin{itemize}
\item radio loudness;
\item rapid variability (high $\Delta L/\Delta t$);
\item high and variable optical polarization ($P_{\rm opt} > 3\%$);
\item smooth, broad, non-thermal continuum; 
\item compact, flat-spectrum radio emission ($f_{\rm core} \gg f_{\rm
extended}$);
\item superluminal motion in sources with multiple-epoch Very Large Baseline
Interferometry (VLBI) maps.  
\end {itemize}

The last property might require some explanation. The term ``superluminal
motion'' describes proper motion of source structure (traditionally mapped at
radio wavelengths) that, when converted to an apparent speed $v_{\rm app}$,
gives $v_{\rm app} > c$. This phenomenon occurs for emitting regions moving at
very high (but still subluminal) speeds at small angles to the line of sight
(Rees 1966). Relativistically moving sources ``run after'' the photons they
emit, strongly reducing the time interval separating any two events in the
observer's frame and giving the impression of faster than light motion. 

Analytically, the observed transverse velocity of an emitting blob, $v_{\rm a}
= \beta_{\rm a} c$, is related to its true velocity, $v = \beta c$, and the
angle to the line of sight $\theta$ by
\begin{equation}
\beta_{\rm a} = {\beta \sin\theta \over 1 - \beta \cos\theta} \quad. 
\label{eq:beta_app}
\end{equation}
It can be shown that if $\beta > 1/\sqrt{2} \simeq 0.7$, then for some
orientations superluminal motion (that is, $\beta_{\rm a} > 1$) is
observed. The maximum value of the apparent velocity, $\beta_{\rm a,max} =
\sqrt{\gamma^2 -1}$, where $\gamma = (1-\beta^2)^{-1/2}$ is the Lorentz
factor, occurs when $\cos \theta = \beta$ or $\sin \theta = \gamma^{-1}$. This
implies a minimum value for the Lorentz factor $\gamma_{\rm min} =
\sqrt{\beta_{\rm a}^2 + 1}$. For example, if one detects superluminal motion
in a source with $\beta_{\rm a} = 5$, the Lorentz factor responsible for it
has to be at least 5.1. 

All these properties are consistent with {\it relativistic beaming}, that is
with bulk relativistic motion of the emitting plasma towards the observer.
There are by now various arguments in favor of relativistic beaming in
blazars, summarized for example by Urry and Padovani (1995). Beaming has
enormous effects on the observed luminosities. Adopting the usual definition
of the relativistic Doppler factor $\delta = [\gamma (1 - \beta \cos
\theta)]^{-1}$
and applying
simple relativistic transformations, it turns out that the {\it observed}
luminosity at a given frequency is related to the {\it emitted} luminosity in
the rest frame of the source via 

\begin{equation}
L_{\rm obs} = \delta^p L_{\rm em} \quad ,
\end{equation}

with $p = 2+\alpha$ or $3+\alpha$ respectively in the case of a continuous jet
or a moving sphere (Urry and Padovani 1995; $\alpha$ being the spectral
index), although other values are also possible (Lind and Blandford 1985).
For $\theta \sim 0^{\circ}$, $\delta \sim 2 \gamma$ (Fig. 2)
and the observed luminosity can be amplified by factors of thousands (for
$\gamma \sim 5$ and $p \sim 3$, which are typical values). That is, for jets
pointing almost towards us we can overestimate the emitted luminosity
typically by three orders of magnitude.  Apart from this amplification,
beaming also gives rise to a strong collimation of the radiation, which is
larger for higher $\gamma$ (Fig. 2): $\delta$ decreases by a
factor $\sim 2$ from its maximum value at $\theta \sim 1/\gamma$ and
consequently the inferred luminosity goes down by $2^p$. For example, if
$\gamma \sim 5$ the luminosity of a jet pointing $\sim 11^{\circ}$ away from
our line of sight is already about an order of magnitude smaller (for $p = 3$)
than that of a jet aiming straight at us.

\begin{figure}
\hskip 2.5truecm 
\psfig{figure=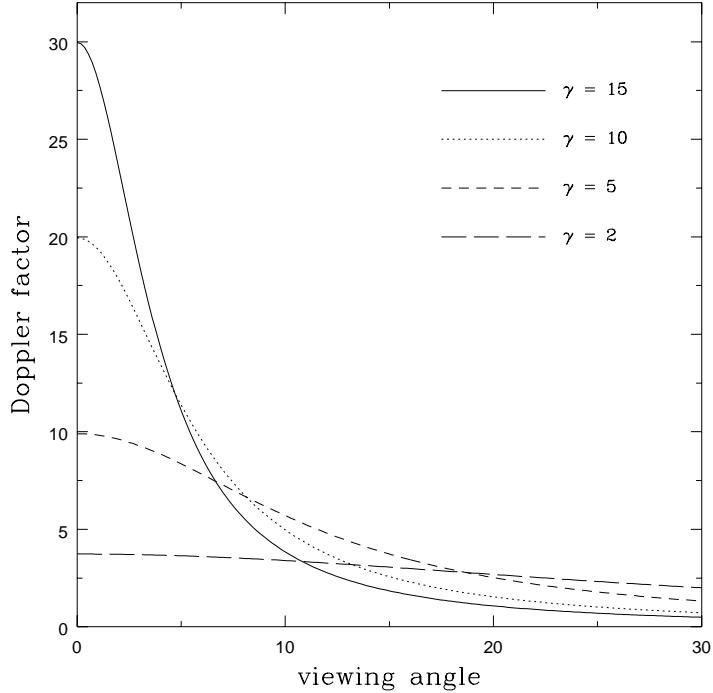,width=9.9truecm}
\caption{The dependence of the Doppler factor $\delta$ on the angle to the 
line of sight. Different curves correspond to different Lorentz factors: from 
the top down, $\gamma = 15$ (solid line), $\gamma = 10$ (dotted line), $\gamma 
= 5$ (short-dashed line), $\gamma = 2$ (long-dashed line).}
\label{fig:doppler}
\end{figure}

All this is very relevant to the issue of $\gamma$-ray emission from blazars.
In fact, if blazars were not beamed, we would not see any $\gamma$-ray photons
from them! The qualitative explanation is relatively simple: in sources as
compact as blazars all $\gamma$-ray photons would be absorbed through
photon-photon collisions with target photons in the X-ray band. The end
product would be electron-positron pairs. But if the radiation is beamed then
the luminosity/radius ratio, which is the relevant parameter, is smaller by a
factor $\delta^{p+1}$ and the $\gamma$-ray photons manage to escape from the
source. More formally, it can be shown (Maraschi et al. 1992) that the
condition that the optical depth to photon-photon absorption $\tau_{\gamma
\gamma}(x)$ is less than 1 implies (under the assumption that the X-ray and
$\gamma$-ray photons are produced in the same region)

\begin{equation}
\delta > C \left({L_{48}\over \Delta t_{\rm d}}\right)^{1/
(p+1)} \left({x \over 10^4}\right)^{\alpha_{\rm x}/(p+1)} \quad ,
\end{equation}

where $L_{48} \equiv L_{\gamma}/(10^{48}$ erg/s), $\Delta t_{\rm d}$ is the
$\gamma$-ray variability time scale in days (which is used to estimate the
source size), $x \equiv h\nu/m_{\rm e} c^2$, $\alpha_{\rm x}$ is the X-ray
spectral index, and $C$ is a numerical constant $\approx 10$. In other words,
transparency for the $\gamma$-ray photons {\it requires} a relatively large
Doppler factor for most blazars (Dondi and Ghisellini 1995) and therefore
relativistic beaming.

\section{The Importance of Being a Blazar}\label{sec:impo}

We can now address the main question of this paper: why have blazars been
detected by EGRET? There are at least three reasons, which have to
do with the fact that blazars are characterized by:

\begin{enumerate}
\item high-energy particles, which can produce GeV photons; 
\item relativistic beaming, to avoid photon-photon collision and amplify the
flux;
\item strong non-thermal (jet) component. 
\end{enumerate}

Point number one is obvious. We know that in blazars synchrotron emission
reaches at least the infrared/optical range, which reveals the presence of
high-energy electrons which can produce $\gamma$-rays via inverse Compton
emission (although hadronic processes can also be important or even dominant;
Mannheim 1993). Very recent BeppoSAX observations have shown that synchrotron
emission can actually reach the X-ray band, namely around 10 keV for 1ES
2344+514 (Giommi, Padovani, and Perlman 1998) and 100 keV for MKN 501 (Pian et
al. 1998). Pian et al. fit a synchrotron-self-Compton (SSC) model to the
multifrequency spectrum of MKN 501 and infer a maximum value for the Lorentz
factor of the electrons $\gamma_{\rm max} \sim 3 \times 10^6$. With such high
values of $\gamma_{\rm max}$, Pian et al. have been able to reproduce the
observed TeV flux of this source (see Sect. \ref{sec:tev}) with an SSC model.
Point number two is vital, as described in the previous section, not only to
enable the $\gamma$-ray photons to escape from the source, but also to amplify
the flux and therefore make the source more easily detectable. Point number
three is also very important. $\gamma$-ray emission is clearly non-thermal
(although we still do not know for sure which processes are responsible for
it) and therefore related to the jet component. The stronger the jet
component, the stronger the $\gamma$-ray flux.

Having understood why blazars have been detected by EGRET, one could also ask:
why have not {\it all} blazars been detected? Many blazars with radio
properties similar to those of the detected sources, in fact, still have only
upper limits in the EGRET band. This problem has been addressed, for example,
by von Montigny et al. (1995) who suggest as possible solutions variability
(only objects flaring in the $\gamma$-ray band can be detected) or a
$\gamma$-ray beaming cone which either points in a different direction or is
more narrowly beamed than the radio one (see also Salamon and Stecker 1994 and
Dermer 1995). These can certainly be viable explanations, but one should also
note that even a moderate dispersion in the values of the parameters required
for $\gamma$-ray emission described above (particle energy, Doppler factor,
and non-thermal component strength) can easily imply the non-detection of some
sources and the detection of others with similar radio properties.

Do the points discussed above also explain why EGRET has not detected any of
the more numerous radio-quiet AGN? Yes, although not all of them might be
essential in this case. As radio-emission (at least in radio-loud AGN) is
certainly non-thermal, while the optical/ultraviolet emission might be thermal
emission associated with the accretion disk (at least in radio-quiet AGN),
then the ratio $L_{\rm r}/L_{\rm opt}$ could actually be related to the ratio
$L_{\rm non-thermal}/L_{\rm thermal}$. Furthermore, $L_{\gamma}$ seems to
scale with $L_{\rm r}$, although the details of this dependence are still
under debate (see e.g., Mattox et al. 1997 and references therein). It
could then be that even radio-quiet AGN are $\gamma$-ray emitters, although
scaled down by their ratio between radio and optical powers, that is at a
level $3 - 4$ orders of magnitude below that of blazars, with typical fluxes
$F_{\gamma} \approx 10^{-10}$ photons cm$^{-2}$ s$^{-1}$. In other words,
radio-quiet AGN would fulfill requirements number one and two\footnote{One
could argue that the relativistic beaming requirement would probably not be
very important in radio-quiet AGN as the luminosity/radius ratio in the
$\gamma$-ray band in these sources would be much lower anyway and $\gamma$-ray
photons would escape even without beaming. However, GeV photons collide
preferentially with X-ray photons, which are plentiful in these sources: some
beaming might then be required for the (putative) $\gamma$-ray emission in
radio-quiet AGN as well.} but not number three. 

Alternatively, it could be that for some reason the emission mechanisms at 
work in radio-loud sources are simply not present in the radio-quiet ones, 
either because there is no jet at all in radio-quiet AGN or because, for
example, there is no accelerating mechanism. In this case, either condition
number one or number three (or both) would be missing (number two would now be
unimportant), and no $\gamma$-ray emission would be expected. 

Unfortunately, it might not be possible to test these two alternatives on the
basis of $\gamma$-ray data for some time: even in the first case, in fact, the
expected $\gamma$-ray fluxes are below the sensitivity of currently planned
future $\gamma$-ray missions, like GLAST (Morselli 1998). If radio-quiet AGN
emit $\gamma$-rays, however, GLAST might be able to detect some of the 
nearest sources. 

\section{AGN as TeV Sources}\label{sec:tev}

So far, only emission up to a few GeV has been considered. However, TeV
astronomy is now in full swing, as we have heard at this meeting, and it is
therefore interesting to consider the situation at these energies. 

Three extragalactic sources have been detected at $E > 0.3$ TeV (Lorenz 1998)
and these are all BL Lacs. That is, even at energies above those of EGRET (and
exactly for the same reasons) the only $\gamma$-ray emitting AGN are still
blazars! The difference here is that, unlike the situation in the EGRET band
where the majority of detected blazars are flat-spectrum radio quasars, only
BL Lacs have been detected so far. Furthermore, these BL Lacs are the three
nearest confirmed BL Lacs in the recent catalogue by Padovani and Giommi
(1995) namely MKN 421 (redshift $z = 0.031$), MKN 501 ($z = 0.055$) and 1ES
2344+514 ($z = 0.044$). The fact that only relatively nearby BL Lacs have been
detected is probably related to absorption of TeV photons by the infrared
background (see, e.g., Biller et al. 1995 and references therein).
Additionally, the position of the synchrotron peak might anticorrelate with
bolometric luminosity in blazars (Fossati et al. 1997). One would then expect
the less luminous, and therefore nearer, objects to have the synchrotron peak
at UV/X-ray energies and the peak of the inverse Compton emission (within the
SSC model) in the TeV band. These objects, therefore, would be, on average,
stronger TeV sources.

Why have no flat-spectrum radio quasars been detected at TeV energies? These
sources are typically at higher redshifts and so the effect on them of the
cosmological absorption by infrared photons is more severe. However, there are
at least four strong radio sources classified as flat-spectrum quasars at $z <
0.1$, including 3C 120 and 3C 111, the latter having been looked at by the
Whipple experiment, with negative results (Kerrick et al. 1995; 3C 111,
however, although superluminal [Vermeulen and Cohen 1994] is 
lobe-dominated [$f_{\rm
core}/f_{\rm extended} \simeq 0.2$; Hes et al. 1995], which suggests it is an
unlikely blazar.  Also, it has not been detected by EGRET.)

This is certainly small number statistics and definite conclusions should only
be drawn after a larger number of relatively local flat-spectrum radio quasars
have been observed at TeV energies. However, the non-detection of
flat-spectrum radio quasars could simply mean that {\it internal} absorption
is significant in these sources. In fact, the cross-section for photon-photon
interaction for $\sim 1$ TeV photons is maximum at $\sim 10^{14}$ Hz or $\sim
2.5 \mu$ and quasars have a larger photon density than BL Lacs at these
frequencies, be it emission from the obscuring torus\footnote{It is not clear
if the presence of an obscuring torus is required in BL Lacs as well as in
radio quasars: see discussion in Urry and Padovani (1995) and Padovani (1997)
and references therein.} or even the accretion disk (see e.g., Protheroe and
Biermann 1997).

The new, more sensitive projects which are underway in the field of TeV
astronomy (e.g., Krennrich 1998) will certainly shed light on these issues and
on the processes which are responsible for the GeV/TeV emission in blazars.
The origin of such emission is in fact still debated as being due to inverse
Compton radiation, either SSC or Comptonization of external radiation (e.g.,
Sikora et al. 1994), or to hadronic processes, that is pion production from
accelerated protons, with subsequent pion decay and $\gamma$-ray production
(e.g., Mannheim 1993). In the latter case, these so-called ``proton blazars''
would also be sources of cosmic rays and of neutrinos and could be detectable
by planned km$^3$ neutrino detectors (Halzen and Zas 1997).

\section{Summary}\label{sec:sum}
The main conclusions are as follows: 

\begin{enumerate}
\item The only AGN detected at GeV and even TeV energies are blazars,
that is a special class of sources which includes BL Lacertae objects and
radio quasars with a relatively flat radio spectrum. These sources are thought
to have their jets moving at relativistic speeds almost directly towards the
observer, a phenomenon which goes under the name of ``relativistic beaming''
and causes strong amplification and collimation of the radiation in our rest
frame.
\item As blazars make up {\it at most} 5\% of all AGN, they must have some
peculiar characteristics which favor their $\gamma$-ray detection. I have
shown that the presence of high-energy particles, relativistic beaming, and a
strong non-thermal (jet) component play, in fact, a fundamental role in making
these sources detectable at $\gamma$-ray energies.
\item $\gamma$-ray missions $\sim 1,000$ times more sensitive than EGRET 
might
also detect the bulk of the more common radio-quiet AGN, under the assumption
that they also possess, on much smaller scales, a non-thermal engine. 
\item TeV astronomy, a very young branch of astronomy which has already
produced some very exciting results, will likely play an important role in the
near future in constraining blazar models. More sensitive TeV telescopes
are clearly needed. 
\end{enumerate}

In summary, there exists a tight connection between unified schemes and 
$\gamma$-ray emission in AGN, as they both depend on relativistic beaming, the
former as a mechanism to produce a strong angle dependence of the observed
properties, the latter as a way to let $\gamma$-ray photons escape from the
source. 


\section{References}

\refe Antonucci, R.: 1993, {\ANNREV} {\bf 31}, p. 473.

\refe Biller, S., et al.: 1995, {\APJ} {\bf 445}, p. 227. 

\refe Dar, A.: 1998, these proceedings. 

\refe Dermer, C.D.: 1995, {\APJL} {\bf 446}, p. 63.  

\refe Dondi, L., Ghisellini, G.: 1995, {\MN} {\bf 273}, p. 583.

\refe Fanaroff, B.L, Riley, J.M.: 1974, {\MN} {\bf 167}, 31p. 

\refe Fossati, G., Celotti, A., Ghisellini, G., Maraschi, L.: 1997, {\MN}
{\bf 289}, p. 136. 

\refe Giommi, P., Padovani, P., Perlman, E.: 1998, {\MN}, submitted. 

\refe Halzen, F., Zas, E.: 1997, {\APJ} {\bf 488}, p. 669. 

\refe Hes, R., Barthel, P.D., Hoekstra, H.: 1995, {\AAP} {\bf 303}, p. 8. 


\refe Kellermann, K.I., Sramek, R., Schmidt, M., Shaffer, D.B., Green, R.: 
1989,  {\AJ} {\bf 98}, p. 1195. 

\refe Kerrick, A.D., et al.: 1995, {\APJ} {\bf 452}, p. 588. 

\refe Krennrich, F.: 1998, these proceedings. 

\refe Lichti, G.G., et al.: 1995, {\AAP} {\bf 298}, p. 711. 

\refe Lind, K.R., Blandford, R.D.: 1985, {\APJ} {\bf 295}, p. 398. 

\refe Lorenz, E.: 1998, these proceedings. 

\refe Mannheim, K.: 1993, {\AAP}, {\bf 269}, p. 67.

\refe Maraschi, L., Ghisellini, G., Celotti, A.: 1992, {\APJL} {\bf 397}, p. 5.

\refe Mastichiadis, A.: 1998, these proceedings. 

\refe Mukherjee, R., et al.: 1997, {\APJ} {\bf 490}, p. 116.  

\refe Mattox, J.R., Schachter, J., Molnar, L., Hartman, R.C., Patnaik, A.R.:
1997, {\APJ} {\bf 481}, p. 95. 


\refe von Montigny, C., et al.: 1995, {\AAP} {\bf 299}, p. 680. 

\refe Morselli, A.: 1998, these proceedings. 


\refe Padovani, P.: 1997, From the Micro- to the Mega-Parsec,  
eds. A. Comastri, T. Venturi, M. Bellazzini, {\SAIT} {\bf 68}, p. 47. 

\refe Padovani, P.: 1998, New Horizons from Multi-Wavelength Sky 
Surveys, eds. B. McLean et al., p. 257.

\refe Padovani, P., Giommi, P.: 1995, {\MN} {\bf 277}, p. 1477. 

\refe Pian, E., et al.: 1998, {\APJL} {\bf 492}, p. 17. 


\refe Protheroe, R.J., Biermann, P.L.: 1997, {\APP} {\bf 6}, p. 293.  


\refe Rees, M.J.: 1966, {\NAT} {\bf 211}, p. 468. 

\refe Salamon, M.H., Stecker, F.W.: 1994, {\APJL} {\bf 430}, p. 21.

\refe Sikora, M., Begelman, M.C., Rees, M.J.: 1994, {\APJ}, {\bf 421}, p. 153.

\refe Stickel, M., Meisenheimer, K., K\"uhr, H.: 1994, {\AAS} {\bf 105}, p. 211.

\refe Stocke, J.T., Morris, S.L., Weymann, R.J., Foltz, C.B.: 1992, 
{\APJ} {\bf 396}, p. 487. 

\refe Thompson, D.J., et al.: 1995, {\APJS} {\bf 101}, p. 259. 

\refe Thompson, D.J., et al.: 1996, {\APJS} {\bf 107}, p. 227. 

\refe Urry, C.M., Padovani, P.: 1995, {\PASP} {\bf 107}, p. 803. 

\refe Vermeulen, R.C., Cohen, M.H.: 1994, {\APJ} {\bf 430} p. 467. 



\end{document}